\documentclass[apj,numberedappendix]{emulateapj}
\usepackage{apjfonts}

\shorttitle{AN ASTEROSEISMIC PIPELINE FOR KEPLER}
\shortauthors{METCALFE ET AL.}

\begin{document}
\def\elll{l}
\def\Sres{S$_\star$}
\def\note #1]{{\bf #1]}}

\title{A Stellar Model-fitting Pipeline for Asteroseismic Data from the Kepler Mission}

\author{T.~S. Metcalfe\altaffilmark{1},
        O.~L. Creevey\altaffilmark{2,3},
        J.~Christensen-Dalsgaard\altaffilmark{4,5}}

\altaffiltext{1}{High Altitude Observatory and Scientific Computing
Division, NCAR, P.O.\ Box 3000, Boulder, CO 80307 USA}
\altaffiltext{2}{Instituto de Astrof{\'\i}sica de Canarias, 38205, La
Laguna, Tenerife, Spain}
\altaffiltext{3}{Newkirk Graduate Fellow, High Altitude Observatory, NCAR}
\altaffiltext{4}{Department of Physics and Astronomy, Aarhus University,
Denmark}
\altaffiltext{5}{Affiliate Scientist, High Altitude Observatory, NCAR}

\slugcomment{Submitted to the Astrophysical Journal (02 March 2009)}

\begin{abstract}

Over the past two decades, helioseismology has revolutionized our 
understanding of the interior structure and dynamics of the Sun. 
Asteroseismology will soon place this knowledge into a broader context by 
providing structural data for hundreds of Sun-like stars. Solar-like 
oscillations have already been detected from the ground in several stars, 
and NASA's Kepler mission is poised to unleash a flood of stellar 
pulsation data. Deriving reliable asteroseismic information from these 
observations demands a significant improvement in our analysis methods. In 
this paper we report the initial results of our efforts to develop an 
objective stellar model-fitting pipeline for asteroseismic data. The 
cornerstone of our automated approach is an optimization method using a 
parallel genetic algorithm. We describe the details of the pipeline and we 
present the initial application to Sun-as-a-star data, yielding an optimal 
model that accurately reproduces the known solar properties.

\end{abstract}



\section{MOTIVATION}

Most of what we can learn about stars comes from observations of their 
outermost surface layers. We are left to infer the properties of the 
interior based on our best current understanding of the constitutive 
physics. The exception to this general rule arises from observations of 
pulsating stars, where seismic waves probe deep through the interior and 
bring information to the surface in the form of light and radial velocity 
variations. The most dramatic example is the Sun, where such observations 
have led to the identification of millions of unique pulsation modes, each 
sampling the solar interior in a slightly different and complementary way. 
The radial profile of the sound speed inferred from inversion of these 
data led to such precise constraints on the standard solar model that, 
before the recent controversy over heavy element abundances 
\citep{asp04,caf08}, the observations and theory agreed to better than a 
few parts per thousand over 90\% of the solar radius \citep{jcd02}. 
Although such detailed inversions are not currently possible for other 
stars, pulsation data do allow us to determine the global properties and 
to probe the gross internal composition and structure, providing valuable 
independent tests of stellar evolution theory. The reason for this 
qualitative difference is simple: if we could move the Sun to the distance 
of the nearest star, many of the pulsation modes would no longer be 
detectable. Without spatial resolution, only those modes with the lowest 
spherical degree ($\elll\la3$) would produce significant variations in the 
total integrated light or the spectral line profiles\footnote{Note that 
pulsation modes of higher spherical degree can sometimes be detected in 
other types of variable stars \citep[e.g.][]{ken98}.}. These are also the 
modes that probe deepest into the stellar interior, collectively sampling 
the physical conditions from the core to the photosphere.

The solar oscillations are excited by the acoustic noise generated by the 
near-surface convection, with nearly sonic speed. Thus we expect to find 
similar oscillations in all stars with vigorous outer convection zones. A 
first unambiguous detection of solar-like oscillations in another star was 
in fact made by \cite{kje95}. Recent improvements in our ability to make 
high-precision radial velocity measurements from the ground have been 
driven largely by efforts to detect extra-solar planets. This has led to 
the detection of solar-like oscillations in more than a dozen 
main-sequence and subgiant stars; in addition, large-scale ground-based 
photometric campaigns have shown evidence for solar-like oscillations in 
numerous giants \citep[for a recent review see][]{bk07}. The oscillation 
amplitudes and the frequency of maximum power in these stars agree 
reasonably well with our theoretical expectations---but in some cases the 
mode lifetimes appear to be significantly shorter than expected 
\citep{ste04}, suggesting that our knowledge of the convective driving and 
damping physics is incomplete. Scintillation in the Earth's atmosphere 
severely limits our ability to detect the parts-per-million light 
variations due to these pulsations in main-sequence stars. However, 
space-based photometric programs designed to detect extra-solar planet 
transits also have the sensitivity to document the stellar pulsation 
signals. Thus, the French-led CoRoT mission \citep{bag06} is providing 
extensive data of very high quality on oscillations in a broad range of 
stars. The field is progressing very rapidly and many new observations 
will become available in the next few years, particularly after the launch 
of NASA's Kepler mission in 2009. With this in mind, it will be a distinct 
advantage to have in place the computational methods that will allow us to 
maximize the science return of these data. This will lead us quickly to a 
deeper understanding of the solar oscillations in the context of similar 
pulsations in other stars, and will yield unprecedented constraints on the 
formation and evolution of stellar and planetary systems.

The likely excitation mechanism by turbulent convection near the surface 
creates a broad envelope of power with a peak that scales with the 
acoustic cutoff frequency \citep[see][]{bg94}. Within this envelope a 
large fraction of the predicted low-degree pulsation modes are excited to 
detectable amplitudes, leading to readily identifiable patterns of peaks 
in the power spectrum characterized by the {\it large separation} 
($\Delta\nu_{\elll} \equiv \nu_{n,\elll} - \nu_{n-1,\elll}$), and the {\it 
small separation} ($\delta\nu_{\elll,\elll+2} \equiv \nu_{n,\elll} - 
\nu_{n-1,\elll+2}$), where $n$ is the radial order and $\elll$ is the 
spherical degree of the modes with oscillation frequency $\nu_{n, \elll}$. 
Without any detailed modeling, these overall patterns immediately lead to 
an estimate of the mean density of the star and can indicate the presence 
of interior chemical gradients that are a good proxy for the stellar age. 
But a full analysis must include a detailed comparison of the individual 
frequencies with theoretical models. One complication with such a 
comparison is the existence of so-called {\it surface effects}, which 
appear as systematic differences between the observed and calculated 
oscillation frequencies that grow larger towards the acoustic cutoff 
frequency. Surface effects arise primarily due to incomplete modeling of 
the near-surface layers of the star where convection plays a major role 
\citep{cdt97}. Addressing this inherent deficiency in our 1D models would 
require (among other things) that we substitute the results of extensive 
3D calculations for the parameterized mixing-length treatment of 
convection that is currently used in nearly all stellar evolution codes. 
Alternately, we can make an empirical correction to the calculated 
frequencies following \cite{kbc08}, who recently devised a method for 
calibrating the functional form of surface effects using solar data, and 
then obtaining corrections for other stars based on the frequencies of a 
reference model, scaled to correct for the different mean density (see 
Appendix~\ref{APPA}).

We have developed an objective and automated method of fitting stellar 
models to the asteroseismic data soon expected to emerge from NASA's 
Kepler mission. This will lead to reliable determinations of stellar radii 
to help characterize the extra-solar planetary systems discovered by the 
mission, and stellar ages to reveal how such systems evolve over time. For 
the asteroseismic targets that do not contain planetary companions it will 
allow a uniform determination of fundamental physical properties for 
hundreds of solar-type stars, providing a new window on stellar structure 
and evolution. In \S\ref{SEC2} we provide an overview of the stellar 
models as well as the optimization and analysis techniques that establish 
the foundation of our fitting method. We describe our initial numerical 
experiments to calibrate the method with Sun-as-a-star data in 
\S\ref{SEC3}, and we outline our plans for further validation in 
\S\ref{SEC4}.


\section{COMPUTATIONAL METHOD\label{SEC2}}

The Kepler mission will soon yield precise high-cadence time-series 
photometry of hundreds of pulsating stars every few months for at least 
3.5 years \citep{jcd07}. We will then face the challenge of determining 
the fundamental properties of these stars from the data, by attempting to 
match them with the output of computer models. The traditional approach to 
this task is to make informed guesses for each of the model parameters, 
and then to adjust them iteratively until an adequate match is found. The 
volume of asteroseismic data that will emerge from the Kepler mission 
calls for a more automated approach to modeling that initially explores a 
broad range of model parameters in an objective manner. The cornerstone of 
our model-fitting approach is a global optimization method using a 
parallel genetic algorithm. The result of the global search provides the 
starting point for a local analysis using a Levenberg-Marquardt algorithm 
with Singular Value Decomposition, which also allows us to explore the 
information content of the observables and the impact of including 
additional observational constraints \citep{bro94,cre07}.

\subsection{Stellar Models\label{sec2.1}}

We have recently adapted the Aarhus stellar evolution code 
\citep[ASTEC;][]{jcd08a} and adiabatic pulsation code 
\citep[ADIPLS;][]{jcd08b} to interface with the parallel genetic 
algorithm. These are essentially the same models that were developed for 
the analysis of helioseismic data, and are the source of Model S of 
\cite{jcd96}, which has been used extensively as a reference model for 
solar inversions. Using these models for the analysis of pulsations in 
solar-type stars will provide some internal consistency in our 
understanding of solar-like oscillations. Briefly, these stellar models 
use the OPAL 2005 equation of state \citep[see][]{rsi96} and the most 
recent OPAL opacities \citep[see][]{ir96}, supplemented by Kurucz 
opacities at low temperatures. The nuclear reaction rates come from 
\cite{bp95}, convection is described by the mixing-length theory of 
\cite{bv58}, and we included the effects of helium settling as described 
by \cite{mp93}.

Each model evaluation involves the computation of a stellar evolution 
track from the zero-age main sequence through a mass-dependent number of 
internal time steps, terminating prior to the beginning of the red-giant 
stage. Rather than calculate the pulsation frequencies for each of the 
200--300 models along the track, we exploit the fact that the average 
frequency spacing of consecutive radial overtones 
$\left<\Delta\nu_0\right>$ in most cases is a monotonically decreasing 
function of age \citep{jcd93}. Once the evolution track is complete, we 
start with a pulsation analysis of the model at the middle time step and 
then use a binary decision tree---comparing the observed and calculated 
values of $\left<\Delta\nu_0\right>$---to select older or younger models 
along the track. In practice, this recipe allows us to interpolate the age 
between the two nearest time steps by running the pulsation code on just 8 
models from each stellar evolution track.

\subsection{Global Search}

Since we are interested in developing a general-purpose modeling tool for 
asteroseismic data from the Kepler mission, we need to select a global 
method for optimizing the match between our model output and the available 
observations of any given star. Using only observations and the 
constitutive physics of the model to restrict the range of possible values 
for each parameter, a genetic algorithm \citep[GA;][]{cha95,mc03} can 
provide a relatively efficient means of searching globally for the optimal 
model. Although it is more difficult for a GA to find {\it precise} values 
for the optimal set of parameters efficiently, it is well suited to search 
for the {\it region} of parameter space that contains the global minimum. 
In this sense, the GA is an objective means of obtaining a good first 
guess for a more traditional local analysis method, which can narrow in on 
the precise values and uncertainties of the optimal model parameters.

\cite{mc03} developed a fully parallel and distributed implementation of 
the PIKAIA genetic algorithm\footnote{The parallel version of the PIKAIA 
genetic algorithm is 
available at http://www.cisl.ucar.edu/css/staff/travis/mpikaia/} that was 
originally written by \cite{cha95}. \cite{mnw00} used this modeling tool 
in the context of white dwarf asteroseismology, which ultimately led to a 
number of interesting physical results, including: (1) a precise estimate 
of the astrophysically important ($^{12}$C + $^4$He $\rightarrow$ 
$^{16}$O) nuclear reaction rate \citep{met03}, (2) the first unambiguous 
detection of a crystallized core in a massive pulsating white dwarf 
\citep{mmk04}, and (3) asteroseismic confirmation of a key prediction of 
diffusion theory in white dwarf envelopes \citep{met05}. The impact of 
this method on the analysis of pulsating white dwarfs suggests that 
seismological modeling of other types of stars could also benefit from 
this approach.

Our implementation of the GA optimizes four adjustable model parameters; 
these are the stellar mass ($M_\star$) from 0.75 to 1.75 $M_\odot$, the 
metallicity ($Z$) from 0.002 to 0.05 (equally spaced in $\log Z$), the 
initial helium mass fraction ($Y_0$) from 0.22 to 0.32, and the 
mixing-length parameter ($\alpha$) from 1 to 3. The stellar age ($\tau$) 
is optimized internally during each model evaluation by matching the 
observed value of $\left<\Delta\nu_0\right>$ (see \S\ref{sec2.1}). The GA 
uses two-digit decimal encoding, so there are 100 possible values for each 
parameter within the ranges specified above. Each run of the GA evolves a 
population of 128 models through 200 generations to find the optimal set 
of parameters, and we execute 4 independent runs with different random 
initialization to ensure that the best model identified is truly the 
global solution. This method requires about $10^5$ model evaluations, 
compared to $10^8$ models for a complete grid at the same sampling 
density, making the GA nearly 1000 times more efficient than a complete 
grid (currently 1 week of computing time, compared to many years for a 
grid). Of course, a grid could in principle be applied to hundreds of 
observational data sets without calculating additional models---but the GA 
approach also gives us the flexibility to improve the physical ingredients 
in the future, while the physics of a grid would be fixed.

\subsection{Local Analysis}

Once the GA brings us close enough to the global solution, we can switch 
to a local optimization method. We implement a modified 
Levenberg-Marquardt (LM) algorithm that uses Singular Value Decomposition 
(SVD) on the calculated design matrices to filter the least important 
information from the observables (some of which may be dominated by 
noise). This then provides an effective local inversion technique. LM is 
relatively fast and stable, and convergence typically occurs within 3--4 
iterations.

We treat the local analysis as a $\chi^2$-minimization problem, where one 
seeks to find the set of parameters {\bf P} that minimizes
\begin{equation}
\chi^2 = \sum_{i=1}^M \left ( \frac{O_i - C_i}{\sigma_i} \right )^2.
\label{eqn:chisq}
\end{equation}
Here $O_i$ and $\sigma_i$ are the $i = 1, 2, \ldots, M$ measurements and 
errors, while the $C_i$ are the calculated model observables resulting 
from {\bf P}. The LM+SVD method requires an initial guess of {\bf P}, and 
these are taken to be the results from the global search by the GA. The 
LM+SVD analysis subsequently uses derivative information from the model at 
the current parameter values to calculate suggested parameter changes 
\boldmath $\delta$\unboldmath {\bf P} that will bring the model 
observables closer to the observations.

We have three main motivations for implementing a local optimization 
method at the end of the global search. First, the GA has a limited 
resolution for each parameter, and the values that match the observations 
best are most likely between the fixed sample points. For example, the 
parameter $Z$ near solar values is tuned by the GA along sample points 
spaced 0.0006 apart, corresponding to a precision of roughly 3\%. 
Considering that the search resolution is limited for {\it all} of the 
model parameters, this significantly limits the precision of the global 
optimization. The resolution of the local analysis is limited only by the 
precision of the stellar evolution and pulsation codes, so we use it to 
adjust the models below the resolution of the GA search.

Our second motivation for the local analysis is to quantify the final 
parameter uncertainties and correlations, and to probe the information 
content of the observables. We do this by calculating the derivatives of 
the model observables with respect to each of the fitted parameters, and 
then dividing each vector by the corresponding measurement error. This 
matrix can be referred to simply as the design matrix {\bf D}. We 
subsequently calculate the singular value decomposition of {\bf D} for 
some of its very useful properties (see Appendix~\ref{APPB}).

The final motivation for implementing a local analysis is to explore the 
effects of using different physical descriptions of the stellar interior 
\citep{cre08a,cre08b}. When the changes to the underlying physics are 
relatively subtle, we can assume that the global search by the GA using 
one set of assumptions will also provide a good starting point for a local 
analysis under slightly perturbed conditions. By applying the techniques 
explained in \cite{cre08b}, such an analysis could demonstrate that the 
observational data contain enough information to distinguish clearly 
between different choices for the equation of state, for example. 
Ultimately, this technique could reveal discrepancies in the observables 
that indicate which physical description is most suitable for the star 
under investigation.

We apply the local analysis to each of the sets of optimal parameters 
identified by the four independent runs of the GA. The local analysis 
involves the gradual refinement of the optimal parameters through the 
iterative application of several steps:
\begin{enumerate}

\item Scan the $Z$ parameter within $\pm$0.002 of its original value, and 
perform a new minimization for each starting value of $Z$. We retain the 
set of parameters {\bf P$_{\rm new}$} that result in the lowest value of 
the reduced $\chi^2$.

\item Starting with {\bf P$_{\rm new}$}, scan the $Y_0$ parameter within 
$\pm$0.01 of its initial value, and perform a new minimization for each 
starting value of $Y_0$. Again we retain the {\bf P$_{\rm new}$} that 
yields the lowest reduced $\chi^2$.

\item Rescan the $Z$ parameter beginning with {\bf P$_{\rm new}$}. We 
adopt as the optimal model the set of parameters that yield the lowest 
reduced $\chi^2$ from this final iteration.

\end{enumerate}
The reason for scanning $Y_0$ and $Z$ becomes clear when we begin to 
understand the intrinsic parameter correlations, which are enhanced by the 
limited set of observations (see \S\ref{sec3.2}).

In most cases the best model found by the global search leads to the final 
best model after the local optimization. However, sometimes a model that 
appears to be marginally worse at the end of the global search is improved 
more substantially during the local analysis. We take our final solution 
to be the best match from the four independent analyses, and for clarity 
we report the results of the corresponding global search even when it is 
not the best of the four models identified by the GA (cf.\ 
Table~\ref{tab1}). The final uncertainties in the parameter values 
($M_\star, Z, Y_0, \alpha, \tau$) and in the model observables ($T_{\rm 
eff}, L_{\star}, R_{\star}$) are calculated using SVD (see 
Appendix~\ref{APPB}).


\section{MODEL-FITTING EXPERIMENTS\label{SEC3}}

The overall goal of our model-fitting pipeline is to take a range of 
oscillation frequencies and other constraints as input, to identify and 
refine the model that best matches these observations, and to produce the 
optimal values of several parameters and other characteristics of the 
model as output. To ensure that our pipeline yields reliable results, we 
must begin by applying it to data where the basic stellar properties are 
already known. The most obvious choice is the Sun, where high-quality 
Sun-as-a-star observations are available from multiple experiments. If we 
feed the pipeline a stellar-like set of solar data, we can judge the 
experiment a success if the pipeline returns the known solar properties 
within acceptable tolerances. To reach this goal, we must first optimize 
the efficiency of the search method by passing synthetic data through the 
model-fitting procedure (\S\ref{sec3.1}), calibrate the differences 
between our models and real observations due to near-surface effects and 
ensure that the resulting empirical correction does not introduce any 
large systematic errors in the final model parameters (\S\ref{sec3.2}), 
and finally quantify any differences in the derived model parameters due 
to the source and error properties of the solar data (\S\ref{sec3.3}). The 
results of these experiments are listed in Table~\ref{tab1}, and described 
in the following subsections. We will eventually want to validate this 
model-fitting pipeline using other stars that differ from the Sun, where 
the physical properties are known with lower precision. However, because 
of the computation-intensive nature of our method (which currently 
requires about 1 week on 512 processors), we will consider here only the 
initial validation using solar data.

\subsection{Efficiency of the Search\label{sec3.1}}

Although genetic algorithms are often more efficient than other comparably 
global optimization methods, they are still quite demanding 
computationally. Fortunately, the procedure is inherently parallelizable; 
we need to calculate many models, and each one of them is independent of 
the others. So the number of available processors determines the number of 
models that can be calculated in parallel. Also, there is very little 
communication overhead; parameter values are sent to each processor, and 
they return either a list of observables or just a goodness-of-fit measure 
if the predictions have already been compared to the observations. The 
parallel version of the PIKAIA genetic algorithm is perfectly general, and 
did not require any structural modifications to interface with our stellar 
evolution and pulsation codes.

The efficiency of genetic-algorithm-based optimization can be defined as 
the number of model evaluations required to yield the global solution, 
relative to the number of models that would be required for a complete 
grid at the same sampling density. In practice, a GA is usually hundreds 
or even thousands of times more efficient than a complete grid, and its 
performance is fairly insensitive to the few internal parameters that 
control its operation. We initially set these internal parameters (e.g.\ 
population size, run length, crossover and mutation rates) based on our 
experience with white dwarf models, but we also ran synthetic data through 
the optimization procedure in a series of ``Hare \& Hound'' (H\&H) 
exercises to ensure that the input parameters were recovered faithfully.

The basic procedure for an H\&H exercise is for one team member to 
calculate the theoretical oscillation frequencies and other observables 
for a specific set of model parameters ($M_\star, Z, Y_0, \alpha, \tau$). 
A subset of the predictions, typical of whatever is available from actual 
observations, is then given to another team member who does not know the 
source parameters. The data are passed through the complete optimization 
method in an attempt to recover the source parameters without any 
additional information. The results of such exercises are used to quantify 
the success rate of the optimization method (the fraction of independent 
runs that lead to the known source parameters), and to improve its 
efficiency (minimize the number of model evaluations) if possible.

For our first H\&H exercise (with source parameters listed in the row 
labeled ``H\&H1'' in Table~\ref{tab1}), we assume that typical 
asteroseismic data from the Kepler mission will include twelve frequencies 
for each of the radial ($\elll=0$), dipole ($\elll=1$), and quadrupole 
($\elll=2$) modes, with consecutive radial orders in the range 
$n=14{-}25$\footnote{Note that only the identification of $\elll$ is 
necessary for the operation of our pipeline. Information about the value 
of $n$ is only provided for completeness.}. Thus, we allowed the GA 
to fit a total of 36 oscillation frequencies. We assigned statistical 
uncertainties to each frequency by scaling up the errors on the 
corresponding modes in BiSON data \citep{cha99} by a factor of 10, which 
is roughly what we expect from Kepler data ($\sigma_\nu\sim0.1~\mu$Hz). We 
complemented this synthetic asteroseismic information with artificial data 
on the effective temperature and luminosity, with errors comparable to 
what is expected for stars in the Kepler Input Catalog \citep[][$T_{\rm 
eff}=5777\pm100$~K, $L_\star/L_\odot=1.00\pm0.1$]{lat05}\footnote{Note 
that the stellar luminosity is a derived quantity based on the observed 
parallax and apparent magnitude.}. This is the simplest conceivable test 
of the pipeline---if we generate a set of model data and then use the 
optimization method to match those data using exactly the same models, how 
quickly will the GA find the true values of the model parameters? The 
results of such a test verify the basic functionality of the algorithm, 
and tell us the approximate number of iterations (or ``generations'' of 
the GA) that we need to execute before stopping the search.

\begin{figure}[t]
\centerline{\includegraphics[angle=0,width=\linewidth]{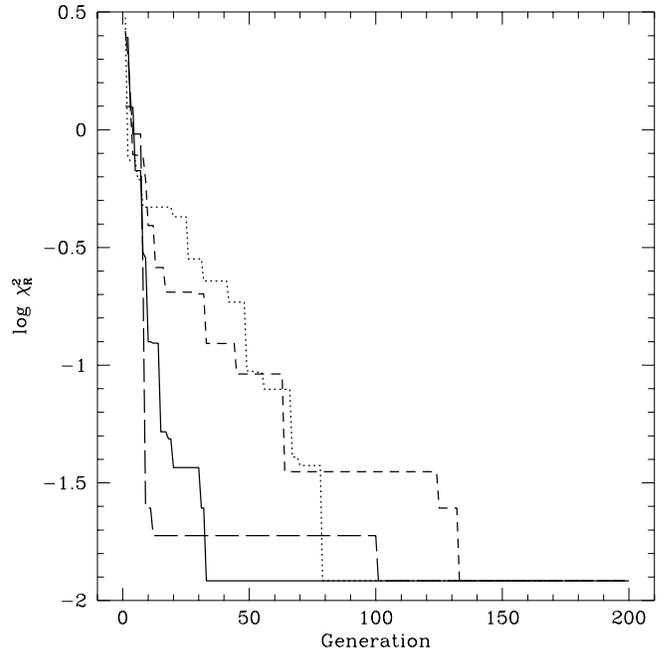}}
\caption{The reduced $\chi^2$ of the best solution in the population as a 
function of generation number for four independent runs of the genetic 
algorithm on the H\&H1 data set. All four runs converge to the global 
solution within 150 generations.\label{fig1}}
\end{figure}

The results of H\&H1 for four independent runs of the GA are illustrated 
in Figure~\ref{fig1}. Each convergence curve shows the reduced $\chi^2$ 
(hereafter $\chi^2_{\rm R}\equiv\chi^2/(M-N-1)$ where $M$ is the number of 
constraints and $N$ is the number of parameters) for the best model in the 
GA population as a function of generation number. All of the runs converge 
in less than $\sim$150 generations, in each case leading to the same mass 
as the source model and values for the other parameters offset by less 
than $\sim$1\% from the source values (the best model from the GA is 
listed in the row labeled ``global'' under H\&H1 in Table~\ref{tab1}). 
Since we expect the algorithm to converge more quickly in this highly 
idealized case, we decided to continue running the GA for 200 generations 
for the more difficult tests that follow. The local analysis using LM+SVD 
brings the final value of the metallicity ($Z$) closer to the source 
model, while retaining comparable accuracy for the other parameters (see 
the row labeled ``local'' under H\&H1 in Table~\ref{tab1}). Note that the 
unusually small values of $\chi^2_{\rm R}$ in Table~\ref{tab1} are a 
consequence of adopting scaled-up BiSON errors for the model frequencies, 
which vastly overestimate the true theoretical uncertainties in this case.

\subsection{Calibrating Surface Effects\label{sec3.2}}

\begin{figure}[t]
\centerline{\includegraphics[angle=270,width=\linewidth]{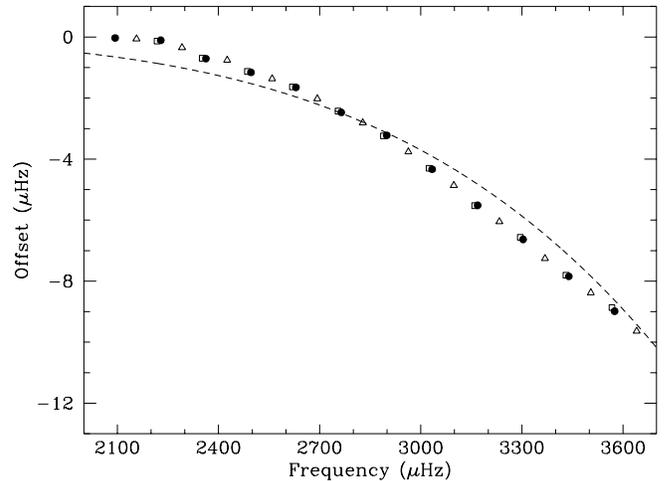}}
\caption{The offset due to surface effects between the Model {\Sres} 
frequencies and the corresponding radial modes (circles) from the BiSON 
data, along with the power law fit to these modes (dashed line). The 
resulting empirical correction is also applied to the dipole (triangles) 
and quadrupole modes (squares), since the offset is not a strong function 
of spherical degree.\label{fig2}}
\end{figure}

The biggest challenge to comparing the oscillation frequencies from 
theoretical models with those actually observed in solar-type stars are 
the systematic errors due to {\it surface effects}. The mixing-length 
parameterization of convection that is used in most stellar models is 
insufficient to describe the near-surface layers, and this leads to a 
systematic difference of several $\mu$Hz (up to about 0.3\% for a solar 
model) between the observed and calculated frequencies (see 
Figure~\ref{fig2}). The offset is nearly independent of the spherical 
degree ($\elll$) of the mode and grows larger towards the acoustic cutoff 
frequency. The 3D simulations of convection that might in principle reduce 
this discrepancy for individual stars are far too computationally 
expensive for the model-fitting approach that we are developing. Instead, 
we adopt the method for empirical correction of surface effects described 
by \cite{kbc08}, which uses the discrepancies between Model S and GOLF 
data for the Sun \citep{laz97} to calibrate the empirical surface 
correction.

The models that we have adopted for the pipeline include a slightly 
different set of physical ingredients than what was used to produce Model 
S from \cite{jcd96}\footnote{Model S also included diffusion and settling 
of heavy elements, and was based on an earlier version of the OPAL 
opacities.}. To characterize the surface effects, we need a model that 
uses the same physics as the other models in the present investigation, 
while matching the Model S frequencies as closely as possible. To find 
such a model (hereafter Model {\Sres}), we used our pipeline to fit the 
computed frequencies of Model S. Since this involves the comparison of two 
sets of model frequencies---both of which include a mixing-length 
parameterization of convection---there are no surface effects to consider. 
Again we allowed the GA to fit 36 frequencies for modes with $n=14{-}25$ 
and $\elll=0{-}2$ and we used the same scaled BiSON errors and the same 
constraints on the effective temperature and luminosity (see 
\S\ref{sec3.1}). The resulting optimal models from the global search and 
the local analysis had the parameter values listed in the rows below 
``Model~{\Sres}'' in Table~\ref{tab1}. Following \cite{kbc08}, we fit a 
power law to the differences between the frequencies of the radial modes 
of this model and the corresponding frequencies from BiSON data to 
characterize the surface effects (see Figure~\ref{fig2} and 
Appendix~\ref{APPA}). We found a power law exponent $b=4.82$, slightly 
lower than the value ($b=4.90$) derived by \citeauthor{kbc08} using data 
from the GOLF experiment. With this exponent fixed, the recipe of 
\citeauthor{kbc08} describes how to predict the surface effect for any 
other set of calculated oscillation data, allowing us to apply this 
empirical correction to each of our models before comparing them to 
observations.

If our strategy of making this empirical correction to each of our models 
is to succeed, it must not only work {\it well} for models in a certain 
region of the search space---it must work {\it best} for the model that 
simultaneously matches all of the independent observational constraints 
within their uncertainties. Our second H\&H exercise (with parameters 
listed in the row labeled ``H\&H2'' in Table~\ref{tab1}) was designed to 
test the behavior of the pipeline with the surface correction included. 
The source model and other constraints were identical to those used for 
H\&H1, but the surface effect was first calculated from 
Eq.(\ref{eq.fobs-fref}) and then applied to each model frequency prior to 
fitting so that the H\&H2 data would mimic real observations. We fixed the 
value of $b$ to 4.82 (as determined above) and we calculated the value of 
$a$ using Eq.(\ref{eq.fobs-fref}) with the scaling factor $r$ determined 
using BiSON data as the reference set of observations. The fitting 
procedure then included the empirical surface correction---attempting to 
remove the systematic frequency errors that we artificially 
introduced---as detailed in Appendix~\ref{APPA}. Again this test reveals 
how long we must run the GA for it to converge to the global solution, but 
it also quantifies any systematic errors on the derived parameter values 
that arise from our implementation of the empirical surface correction.

As expected, the GA converges to the global solution more slowly with the 
inclusion of the surface correction. Within 150--200 generations, two of 
the four runs converged to the parameters listed in the row labeled 
``global'' under H\&H2 in Table~\ref{tab1}, while the other two remained 
in a nearby local minimum. If we had continued the search, these two runs 
would eventually also have found the global solution. The largest 
systematic error appears in the value of $Z$, which is about 6\% (two 
sample points) higher in the global solution than in the source model. 
There are also smaller errors ($\sim$1\%) on the derived values of $Y_0$ 
and $\tau$. The results of the local analysis (listed in the row labeled 
``local'' under H\&H2 in Table~\ref{tab1}) bring both $Z$ and $\tau$ 
closer to the source values while preserving the accuracy of the other 
parameters. The remaining errors are not a deficiency in the ability of 
the method to find the true solution---the identified model actually 
results in a lower $\chi^2_{\rm R}$ than the source parameters.

The reason for this counter-intuitive result is that we can only generate 
a surface correction when there is a reference set of observed oscillation 
frequencies (see Appendix~\ref{APPA}). Our source model used the BiSON 
data for reference, and these data have a slightly different value of the 
large frequency separation ($\left<\Delta\nu_0\right>=134.673\,\mu$Hz) 
than the resulting source model 
($\left<\Delta\nu_0\right>=134.759\,\mu$Hz). Since this quantity is used 
by the binary decision tree to fit the age of the model, the small 
difference leads to a slightly offset derived age---which subsequently 
modifies the optimal composition through intrinsic parameter correlations. 
In other words, when generating a model with the surface correction 
included from a given set of parameters, those same parameters will 
generally not provide the best fit to the resulting observables. This 
actually highlights our inability to generate realistic data with surface 
effects, rather than an inherent limitation in our fitting method. In any 
case, this exercise demonstrates that the systematic errors resulting from 
our surface correction are small, and that running the GA for 200 
generations should be sufficient for real observations.

\subsection{Validation with Solar Data\label{sec3.3}}

Ultimately, our model-fitting pipeline can only be judged a success if it 
leads to accurate estimates of the stellar properties for the star that we 
know best: the Sun. Up to this point, we have essentially been fitting 
models to synthetic data---using the solar data from BiSON only to 
calibrate the empirical surface correction and to provide realistic 
errors. There are many other ingredients in our models that could in 
principle be insufficient descriptions of the actual conditions inside of 
real stars---deficiencies that could easily lead to systematic errors in 
our determinations of the optimal model parameters for a given set of 
oscillation data. For example, we initially tried to use models that 
employed the simpler EFF equation of state \citep{eff73} for computational 
expediency, but this led to estimates of the stellar mass about 10\% too 
high for the Sun, and unacceptably large systematic errors on many of the 
other stellar properties. Even attempting to ignore the effects of helium 
settling proved to be too coarse an approximation, leading to 5\% errors 
on the mass. The only potential ingredient that we omitted without serious 
consequences was heavy element settling. This is not to say that simpler 
stellar models cannot be used in the analysis of asteroseismic data, but 
rather that some of the more sophisticated ingredients are required to 
obtain accurate results from a {\it global} search of the parameter space.
 
Having demonstrated the effectiveness of the method by fitting our models 
to synthetic data, and after calibrating the empirical surface correction 
using the differences between Model {\Sres} and the BiSON data, we finally 
applied our model-fitting pipeline to solar data from the BiSON and GOLF 
experiments. The oscillation frequencies from these two sources are 
identical to each other within the observational uncertainties, but their 
noise properties are slightly different---allowing us to quantify any 
systematic errors that might arise from subtle effects in the data 
acquisition and analysis methods. In both cases we used the same set of 
modes referenced in the earlier experiments ($n=14{-}25$, $\elll=0{-}2$) 
with the respective errors again scaled up by a factor of 10, and the same 
constraints on the effective temperature and luminosity (see 
\S\ref{sec3.1}). The two sets of input data differed only in the absolute 
values of the oscillation frequencies (yielding distinct values of 
$\left<\Delta\nu_0\right>$ for fitting the stellar age), and in the 
statistical uncertainties assigned to each mode (leading to subtle 
differences in the weighting of the fit).

\begin{figure}[t]
\centerline{\includegraphics[angle=0,width=\linewidth]{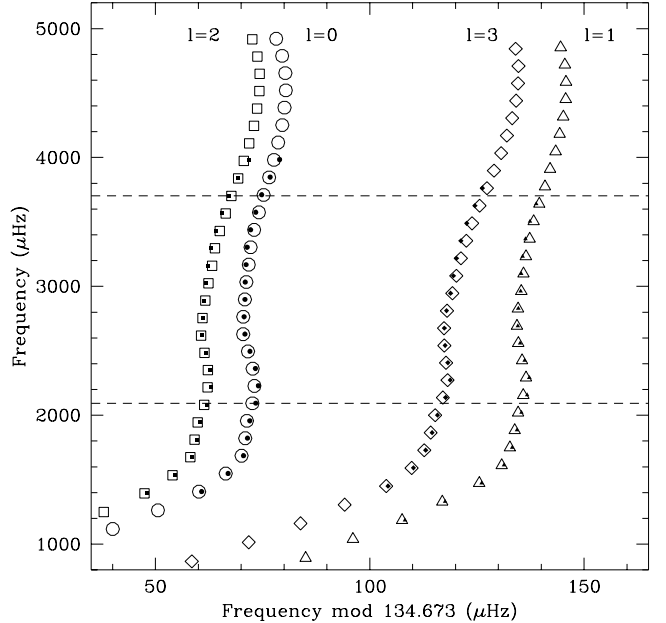}} 
\caption{An echelle diagram for the BiSON data (solid points), where we 
divide the oscillation spectrum into segments of length 
$\left<\Delta\nu_0\right>$ and plot them against the oscillation 
frequency, along with the optimal model from our asteroseismic modeling 
pipeline (open points). Note that the pipeline only used the $\elll=0{-}2$ 
frequencies between the dashed lines for the fit, but the resulting 
optimal model also matches the $\elll$=3 modes and frequencies outside of 
the fitting range.\label{fig3}}
\end{figure}

In general, the four independent runs of the GA for each data set led to 
slightly different results after 200 generations, so we list only the 
solution leading to the best model in the rows below ``BiSON'' and 
``GOLF'' in Table~\ref{tab1}. We found that if we continued running the GA 
for up to 300 generations, the same optimal solution was identified in 
2--3 of the independent runs. To reduce the total computing time required 
for future experiments, we decided to stop the GA earlier when 1--2 of the 
runs would still reliably identify the global solution. Both data sets 
lead to identical values of the mass and metallicity from the global 
search, with slight variations in the values of the other parameters. 
These minor differences largely disappear after the local analysis. Note 
that because we multiplied the true observational errors by a factor of 10 
for the fitting, the resulting values of $\chi^2_{\rm R}$ are $\sim$0.1. 
Although the fits used a limited range of frequencies and did not include 
$\elll=3$ modes, the optimal models also match the modes with lower 
frequencies and higher degree (see the BiSON fit in Figure~\ref{fig3}) and 
accurately reproduce the known solar properties.

\tablewidth{500pt}
\tabletypesize{\normalsize}
\tablecaption{Results of Model-fitting Experiments\label{tab1}}
\begin{deluxetable*}{rccccccccc}
\tablehead{
\colhead{Run}               &
\colhead{$M_\star~(M_{\odot})$} &
\colhead{$Z$}               &
\colhead{$Y_0$}             &
\colhead{$\alpha$}          &
\colhead{$\tau~(Gyr)$}      &
\colhead{$T_{\rm eff}~(K)$} &
\colhead{$L_\star~(L_{\odot})$} &
\colhead{$R_\star~(R_{\odot})$} &
\colhead{$\chi^2_{\rm R}$} }
\startdata
H\&H1   & 1.000 & 0.0191 & 0.271 & 2.06 & 4.54 & 5780 & 1.001 & 1.000 &$\cdots$\\
global: & 1.000 & 0.0197 & 0.273 & 2.04 & 4.51 & 5763 & 0.990 & 1.000 & 0.052 \\
local:  & 1.001 & 0.0193 & 0.269 & 2.03 & 4.58 & 5755 & 0.986 & 1.000 & 0.004 \\
error:  & 0.009 & 0.0014 & 0.012 & 0.07 & 0.16 &~~~61 & 0.040 & 0.004 &$\cdots$\\
\hline
Model~{\Sres}&$\cdots$&$\cdots$&$\cdots$&$\cdots$&$\cdots$&$\cdots$&$\cdots$&$\cdots$&$\cdots$\\
global: & 0.990 & 0.0191 & 0.277 & 2.04 & 4.56 & 5778 & 0.993 & 0.996 & 0.062 \\
local:  & 0.993 & 0.0188 & 0.275 & 2.05 & 4.52 & 5787 & 1.001 & 0.997 & 0.039 \\
error:  & 0.009 & 0.0013 & 0.010 & 0.07 & 0.16 &~~~60 & 0.042 & 0.003 &$\cdots$\\
\hline
H\&H2   & 1.000 & 0.0191 & 0.271 & 2.06 & 4.54 & 5780 & 1.001 & 1.000 &$\cdots$\\
global: & 1.010 & 0.0210 & 0.272 & 2.12 & 4.67 & 5773 & 1.005 & 1.004 & 0.042 \\
local:  & 1.010 & 0.0197 & 0.268 & 2.09 & 4.55 & 5776 & 1.006 & 1.003 & 0.007 \\
error:  & 0.010 & 0.0016 & 0.012 & 0.07 & 0.16 &~~~61 & 0.041 & 0.004 &$\cdots$\\
\hline
BiSON   &$\cdots$&$\cdots$&$\cdots$&$\cdots$&$\cdots$&$\cdots$&$\cdots$&$\cdots$&$\cdots$\\
global: & 1.010 & 0.0217 & 0.278 & 2.14 & 4.52 & 5793 & 1.018 & 1.004 & 0.197 \\
local:  & 1.012 & 0.0200 & 0.267 & 2.12 & 4.65 & 5779 & 1.010 & 1.004 & 0.146 \\
error:  & 0.008 & 0.0014 & 0.010 & 0.06 & 0.17 &~~~55 & 0.039 & 0.003 &$\cdots$\\
\hline
GOLF    &$\cdots$&$\cdots$&$\cdots$&$\cdots$&$\cdots$&$\cdots$&$\cdots$&$\cdots$&$\cdots$\\
global: & 1.010 & 0.0217 & 0.273 & 2.10 & 4.71 & 5752 & 0.990 & 1.004 & 0.183 \\
local:  & 1.011 & 0.0199 & 0.267 & 2.11 & 4.66 & 5779 & 1.008 & 1.004 & 0.124 \\
error:  & 0.014 & 0.0017 & 0.017 & 0.07 & 0.21 &~~~67 & 0.043 & 0.005 &$\cdots$
\enddata
\end{deluxetable*}

\section{DISCUSSION\label{SEC4}}

With the successful validation of our model-fitting pipeline using solar 
data, we now need to ensure that our adopted treatment of surface effects 
yields reasonable optimal models when applied to other stellar data. 
While numerous ground-based and space-based data sets on other stars are
already available, the computation-intensive nature of our method
necessarily limits the present investigation to the single case of the
Sun. The obvious next step is to use archival ground-based data on several 
well studied solar-type stars to validate the pipeline for various stellar 
masses (e.g. $\alpha$ Cen A \& B, near $\sim$1.1 and $\sim$0.9 $M_\odot$ 
respectively) and for different evolutionary stages (e.g. the ``future 
Sun'' $\beta$~Hyi, at $\sim$7~Gyr). Although \cite{kbc08} demonstrated 
their method by applying it to models of these three stars in addition to 
the Sun, our experience using it with solar data suggests that a {\it 
global} exploration of the models may present additional challenges.

Since our stellar models and the empirical correction for surface effects 
have both been calibrated using a main-sequence star at 1.0~$M_\odot$, the 
$\alpha$~Cen system will help validate the models with interior physical 
conditions that differ slightly from those of the Sun. Its proximity and 
multiple nature make it an excellent second test of our pipeline, because 
it has very well determined properties including stellar radii from 
interferometry \citep{ker03}. There are also strong constraints on the 
component metallicities and effective temperatures \citep{plk08}, while 
the initial composition and age of the two stars are presumably identical. 
In the next phase of this project, we plan to use the published 
oscillation frequencies of $\alpha$~Cen~A \citep{bc02,bed04,baz07} and 
$\alpha$~Cen~B \citep{cb03,kje05} with the additional constraints from 
interferometry, spectroscopy, and the binary nature of the system to 
further validate our pipeline and the empirical correction for surface 
effects. \cite{kbc08} successfully applied their recipe to a set of 
stellar models that broadly resemble the components of the $\alpha$~Cen 
system, so we have good reason to believe that our implementation will 
also succeed---but this remains to be demonstrated.

The G2 subgiant $\beta$~Hyi has long been studied as a ``future Sun'', 
with an age near 7~Gyr. It has been characterized almost as extensively as 
the $\alpha$~Cen system, including recent interferometric measurements of 
its diameter \citep{nor07} and dual-site asteroseismic observations 
that determined its mean density with an accuracy of 0.6\% \citep{bed07}. 
These data included the detection of several $\elll=1$ modes that deviate 
from the asymptotic frequency spacing, suggesting that they are ``mixed 
modes'' behaving like g-modes in the core and p-modes in the envelope. 
This is expected for evolved stars like $\beta$~Hyi because as they expand 
and cool the p-mode frequencies decrease, while the g-mode frequencies 
increase as the star becomes more centrally condensed. This leads to a 
range of frequencies where these modes can overlap and exchange their 
character, manifested as so-called {\it avoided crossings}. This behavior 
changes very quickly with stellar age, and propagates from one mode to the 
next as a star continues to evolve. Consequently, the particular mode 
affected yields a very strong constraint on the age of the star 
\citep[see][]{jcd04}. In subsequent work, we plan to use the published 
oscillation data for $\beta$~Hyi \citep{bed07} along with the constraints 
from interferometry and spectroscopy to validate our pipeline and the 
empirical treatment of surface effects for stars that are significantly 
more evolved than the Sun. This will require an automated method to 
recognize mixed modes in the data set and to incorporate them into the 
optimization of stellar age along each track.

Once we have validated the model-fitting pipeline with additional stars 
that sample a range of masses and evolutionary stages, we can begin to 
consider additional observables and parameters that are not constrained by 
currently available data sets. High-quality asteroseismic data are soon 
expected from the Kepler mission, spanning sufficiently long periods of 
time that the effects of rotation \citep{gs03,bal06,bal08} and magnetic 
activity cycles \citep{cha07,met07} should be detectable. The Kepler 
mission is designed to discover Earth-sized habitable planets, and our 
model-fitting pipeline will be able to characterize the planet-hosting 
stars with asteroseismology. This is essential to convert precise transit 
photometry into an absolute radius for the planetary body. In addition, 
accurate rotation rates and ages will provide clues about the formation 
and evolution of the planet-hosting systems. The determination of accurate 
stellar properties for a broad array of solar-type stars will give us a 
new window on stellar structure and evolution, and will provide a broader 
context for our understanding of the Sun and our own solar system. We hope 
to facilitate this process by applying our stellar model-fitting pipeline 
to the data that will soon emerge from the Kepler mission.

\acknowledgments
The authors wish to thank Tim Brown and Margarida Cunha for helpful 
discussions during the early phases of this project. This work was 
supported in part by an NSF Astronomy \& Astrophysics Fellowship (to 
T.S.M.) under award AST-0401441, by a Newkirk Graduate Fellowship (to 
O.L.C.) at the High Altitude Observatory and by the European Helio- and 
Asteroseismology Network (HELAS) a major international collaboration 
funded by the European Commission's Sixth Framework Programme, by the 
Danish Natural Science Research Council, and by NASA grant NNX09AE59G. 
Computer time was provided by NSF MRI grants CNS-0421498, CNS-0420873, 
CNS-0420985, the University of Colorado, and a grant from the IBM Shared 
University Research (SUR) program. The National Center for Atmospheric 
Research is a federally funded research and development center sponsored 
by the U.S.~National Science Foundation.


\appendix

\section{THE EMPIRICAL SURFACE CORRECTION\label{APPA}}

For convenience we summarize the analysis of \citet{kbc08}, to correct for 
the near-surface frequency effects. This is based on analyzing the large 
frequency separation $\Delta \nu_{n,\elll}$ (cf.\ \S 1) and the individual 
frequencies. Since the offset from incorrect modeling of the near-surface 
layers is not a strong function of~$\elll$, we can derive the correction 
by considering only the radial modes ($\elll=0$) and then apply it to all 
of the modes.

Suppose we have a set of observed frequencies for radial modes $\nu_{n, 
0}^{\rm (obs)}$, where $n$ is the radial order. Let $\nu_{n, 0}^{\rm 
(best)}$ be the frequencies of the model that we seek---the one that best 
describes the parameters and internal structure of the star, but which 
still fails to model correctly the surface layers. In the solar case, this 
model is defined by the inversion of the frequencies over a broad range of 
available modes. Here the difference between the observed and best model 
frequencies is found to be well fitted by a power law, which has the 
convenient property of being free of a frequency scale:
\begin{equation}
  \nu_{n, 0}^{\rm (obs)} - \nu_{n, 0}^{\rm (best)} =
a \left(\frac{\nu_{n, 0}^{\rm (obs)}}{\nu_0}\right)^b,
\label{eq.fobs-fref}
\end{equation}
\citep[see also][]{jcd80}, where $\nu_0$ is a suitably chosen reference 
frequency, and $a$ and $b$ are parameters to be determined.

In practice we do not know the best model. Thus, we must determine the 
correction from a reference model, with frequencies $\nu_{n, 0}^{\rm 
(ref)}$, which is assumed to be close to the best model. In that case we 
have to a good approximation from homology scaling
\begin{equation}
  \nu_{n, 0}^{\rm (best)} = r \nu_{n, 0}^{\rm (ref)},\label{eq.fbest.fref}
\end{equation}
where the scaling factor $r$ is related to the mean densities of the best 
and reference models.

Substituting Eq.(\ref{eq.fbest.fref}) into Eq.(\ref{eq.fobs-fref}) and 
differentiating with respect to $n$ gives
\begin{equation}
  \Delta\nu_{n, 0}^{\rm (obs)} - r \Delta\nu_{n, 0}^{\rm (ref)} = 
a b \left(\frac{\nu_{n, 0}^{\rm (obs)}}{\nu_0}\right)^{b-1} 
\frac{\Delta\nu_{n, 0}^{\rm (obs)}}{\nu_0},
\label{eq.Deltanu}
\end{equation}
from which we finally obtain
\begin{equation}
  r = (b-1)\left(b \frac{\nu_{n, 0}^{\rm (ref)}}{\nu_{n, 0}^{\rm (obs)}} 
- \frac{\Delta\nu_{n, 0}^{\rm (ref)}}{\Delta\nu_{n, 0}^{\rm (obs)}}\right)^{-1} 
\label{eq.r}
\end{equation}
and
\begin{equation}
  b = \left(r \frac{\Delta\nu_{n, 0}^{\rm (ref)}}{\Delta\nu_{n, 0}^{\rm (obs)}} - 1\right) 
\left(r \frac{\nu_{n, 0}^{\rm (ref)}}{\nu_{n, 0}^{\rm (obs)}} - 1\right)^{-1}
\label{eq.b}
\end{equation}
\citep[for details, see][]{kbc08}. If we know $b$ then we can calculate 
$r$ using Eq.(\ref{eq.r}), or {\em vice versa\/} using Eq.(\ref{eq.b}). We 
can then obtain~$a$ using equations~(\ref{eq.fobs-fref}) and 
(\ref{eq.fbest.fref}).

Applying this method to the Sun, we used the frequencies of Model~{\Sres}, 
which were obtained by fitting the frequencies of Model S from 
\citet{jcd96}, but using slightly modified physics (see \S\ref{sec3.2}). 
We assumed this to be the ``best'' solar model, which means that we can 
set $r=1$ (see Eq.~\ref{eq.fbest.fref}). The observed solar frequencies 
were obtained from the BiSON experiment \citep{cha99}. From the procedure 
described above, choosing $\nu_0 = 2832.82\,\mu$Hz, setting $r=1$ and 
using the data for the dozen radial modes with $n=14{-}25$, we obtained a 
value of $b=4.82$.

\cite{kbc08} repeated a similar analysis for different numbers of modes 
(all values from 7 to 13), and found the derived value of $b$ to range 
from 4.4 to 5.25. Clearly, the frequency differences do not exactly follow 
a power law, and so the exponent in the power-law fit depends 
substantially on the frequency range. However, the value for $a$ varied by 
less than 0.1\,$\mu$Hz in all cases. If we repeat our analysis using GOLF 
data \citep{laz97} for the same set of modes, we derive a value of 
$b=4.97$. Both of our values are comparable to the value $b=4.90$ derived 
by \cite{kbc08}.

After calibrating the value of the exponent $b$ using solar data, the 
prescription of \citeauthor{kbc08} describes how to predict the surface 
effect for any other stellar model being compared to real observations. 
The value of $r$ is determined by applying Eq.(\ref{eq.r}) to each trial 
model through a comparison of the calculated frequencies with the 
observations. The value of $a$ is then determined using equations 
(\ref{eq.fobs-fref}) and (\ref{eq.fbest.fref}), and subsequently used in 
conjunction with $b$ to predict the surface effect for each frequency of 
the model. This recipe is applied to every model calculated during the 
global search by the GA, and during the local analysis using LM+SVD. 
Effectively we are adjusting the predicted surface effect for each model 
based on the difference between the mean densities implied by the two sets 
of frequencies. In addition, the magnitude of the surface effect reduces 
the weight of the mode in the final fit, giving more weight to low 
frequencies but still attempting to optimize the match at high 
frequencies. The final weight for each mode is determined by the quadratic 
sum of the (statistical) observational error and half the (systematic) 
surface correction.


\section{SINGULAR VALUE DECOMPOSITION\label{APPB}}

As discussed in detail by \citet{bro94}, the analysis of the linearized 
relation between observables and parameters is conveniently carried out in 
terms of Singular Value Decomposition (SVD) because it has some very 
useful properties when applied to a matrix. Suppose we have a {\it model} 
that when given a set of input parameters {\bf P} (we normally do not know 
{\bf P} and we wish to retrieve them) produces an output set of expected 
model observables. Generally the model- (or parameter-) fitting problem 
consists of matching a set of observations {\bf O} to the output model 
observables, by making parameter changes \boldmath $\delta$\unboldmath 
{\bf P} to {\bf P}, until we find the model that fits the observations 
best. Once we find {\bf P}, to exploit SVD we calculate the partial 
derivatives of each of the observables with respect to each of the model 
parameters {\bf P}, and subsequently divide by the measurement errors. The 
result is what we call the design matrix {\bf D}.

The SVD of {\bf D} (= {\bf UWV$^{\rm T}$}) neatly describes the 
relationship between the model observables and the stellar parameters 
through a set of linear transformation vectors. Here the matrices {\bf U} 
and {\bf V} are of order $M \times N$ and $N \times N$ respectively, where 
$M$ is the number of observables and $N$ is the number of parameters; {\bf 
W} is a diagonal matrix of order $N \times N$, with diagonal elements 
consisting of the singular values $W_k$. {\bf U} and {\bf V} consist of 
orthonormal vectors that span the observable and parameter spaces 
respectively, while the diagonal elements of {\bf W} assign a relative 
importance to each vector. Each column vector of {\bf U} describes the 
content of each observable, the singular value $W_k$ quantifies the 
importance of this vector (i.e., it organizes the observables by amount 
and type of information), while each column vector of {\bf V} describes 
how the information from each observable vector is distributed among the 
parameters (i.e., which observables are important for constraining certain 
parameters). We will see this more clearly below. If the models were 
linear then the solution \boldmath $\delta$\unboldmath {\bf P}, in a 
least-squares sense, could be found simply from the discrepancies 
\boldmath $\delta$\unboldmath {\bf O} between the observations and the 
model observables, normalized by the errors. In terms of SVD, the result 
can be written as \boldmath $\delta$\unboldmath {\bf P} = {\bf 
VW$^{-1}$U$^{\rm T}$} \boldmath $\delta$\unboldmath {\bf O}, which shows 
clearly that the parameter changes required to reconcile {\bf P} with the 
observables comes from a product of (1) the discrepancies between the 
observations and the observables, and (2) a set of linear transformation 
vectors that are scaled by the inverse of the singular values. Below we 
will explore how we can use SVD to illucidate both the information 
contained in the set of observables and the available constraints on the 
parameters.

\begin{figure}[t]
\centerline{\includegraphics[angle=0,width=5.5in]{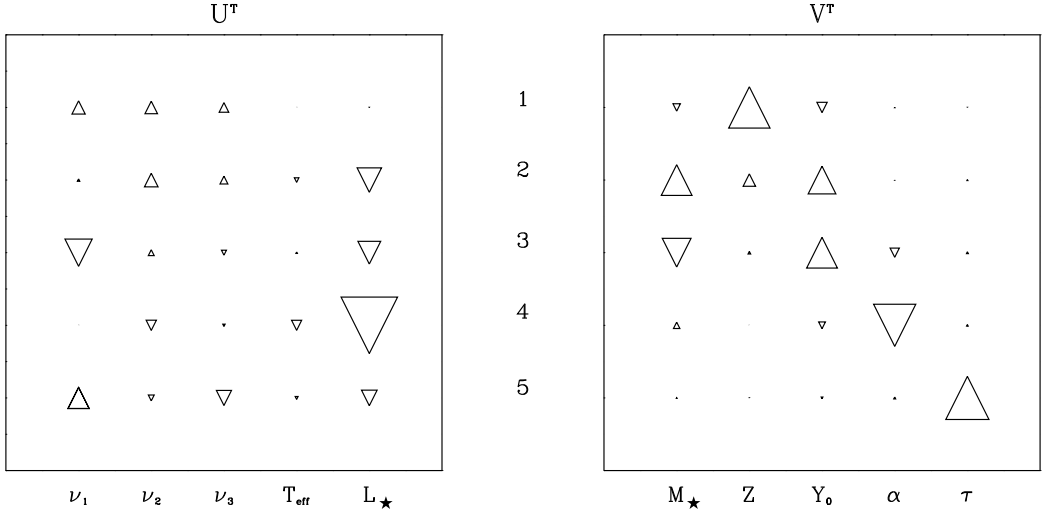}}
\caption{The ${\bf U}^{\rm T}$ and ${\bf V}^{\rm T}$ orthonormal matrices 
from the SVD of the design matrix, evaluated for the model that best fits 
the BiSON data. Each row in ${\bf U}^{\rm T}$ shows how the set of 
observables are responsible for each of the parameters given in each row 
of ${\bf V}^{\rm T}$, e.g., row 1 contains mostly the parameter of $Z$, 
and this is constrained mainly by the 3 frequencies that appear in row 1 
of ${\bf U}^{\rm T}$. If we scale the rows in ${\bf V}^{\rm T}$ by the 
inverse of the singular values we obtain the parameter correlations, which 
describe how to define the 5-dimensional error boundary.\label{fig4}}
\end{figure}

Figure \ref{fig4} illustrates part of the orthonormal matrix {\bf U} 
(left) and the full orthonormal matrix {\bf V}.  Because we have $N = 5$ 
parameters in the problem, each of the matrices consists of 5 {\it column 
vectors};  thus we plot ${\bf U}^{\rm T}$ and ${\bf V}^{\rm T}$, ordered 
such that the horizontal vectors corresponding to the largest singular 
values $W_k$ are at the top. Each column vector of {\bf U} has $M = 38$ 
elements, which is the number of observations that are used to determine 
the solution, but we only include the first 5 elements for illustration. 
The elements of the matrix can only take values between $-1$ and +1, and 
each number is represented by a triangle whose size and direction are 
proportional to the magnitude and sign of each element. The column vectors 
are labeled 1--5: 1 represents the first column vector, which is 
associated with the largest singular value, 2 is the second largest 
singular value, while 5 is the smallest.  Thus, the vectors are 
effectively {\it ranked} according to their information content, where the 
least important information is given in the vectors corresponding to the 
smallest $W_k$. For example, if we were to perform the matrix 
multiplication of {\bf UWV$^{\rm T}$} using only the top vectors, we would 
be able to recover most of the input matrix {\bf D}. Essentially this is 
equivalent to zeroing the values of {\bf W} that are below a certain 
threshold before doing the multiplication. Clearly the observables that 
appear among the top vectors are most important for constraining the 
parameter solution, while the parameters that appear among the top vectors 
are the most constrained.

Inspecting the left panel of Figure~\ref{fig4}, by scanning down the 
fourth column we see that the observable $T_{\rm eff}$ has components 
mainly in the second and fourth vectors. The luminosity $L_\star$ is 
present in vectors 2--4, while the individual frequencies are present 
mostly in vectors 1 and 2. The most important information comes from the 
first vector, and we can see clearly that in this case the seismic 
information dominates. We also see that $L_\star$ and $T_{\rm eff}$ 
usually appear in the same vectors, implying that these observables work 
together to constrain the parameter solution. The right panel shows how 
the information from the observables is distributed among the parameters.  
The top vector has mainly one large component, $Z$.  This implies that $Z$ 
is most constrained by the set of observables, and in particular by the 
observables that appear in the top row of the ${\bf U}^{\rm T}$ matrix, 
i.e. the frequencies. Similarly, the mixing-length parameter ($\alpha$) 
appears predominantly in the fourth vector, and is constrained mainly by 
$T_{\rm eff}$ and $L_\star$.

Returning to the linear parameter-fitting problem, we saw that calculating 
\boldmath $\delta$\unboldmath {\bf P} depended strongly on the size of the 
singular value associated with the transformation vectors (as well as the 
discrepancies).  Because $Z$ appears in the first vector and the inverse 
of the first singular value is the smallest, this means that $Z$ is not 
allowed to deviate much from its initial value. So if we find that we are 
in a local minimum that is not the global solution, it is unlikely that we 
will find a better solution that deviates much in $Z$. For this reason we 
choose to explore a range of $Z$ values during the local analysis. The 
vectors {\bf V}$_{\bf k}$ are the basis vectors, and describe the 
normalized changes that we can make to each of the parameters while 
respecting the constraints imposed by the observables. The inverse of the 
$W_k$ describe the magnitude of each of the vectors that keeps the 
parameters within 1$\sigma$ of the final answer. Since the vectors 
corresponding to larger $k$ have smaller $W_k$ (larger inverses), the 
parameters associated with these vectors will extend further into the 
parameter space, i.e. they are less constrained by the observations.

\begin{figure}[t]
\centerline{\includegraphics[angle=0,width=3.5in]{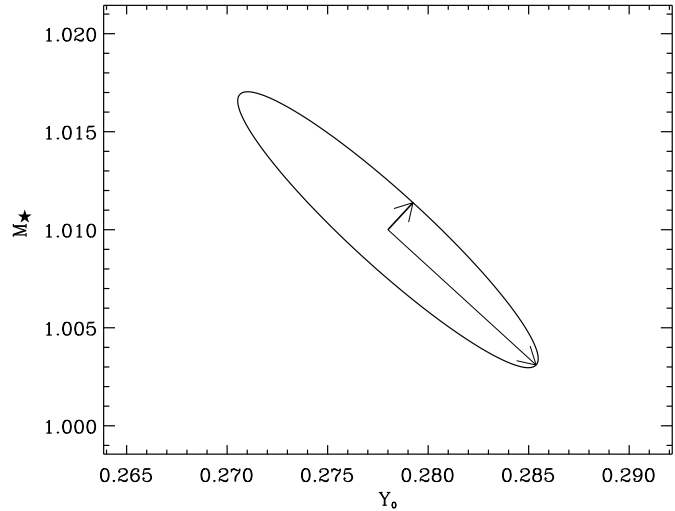}}
\caption{The longest projections of the 5-dimensional error boundary on 
the 2-dimensional parameter space of $M_\star$ and $Y_0$, as described by 
SVD, for the final fit model to the BiSON data. The arrows are the 
2-element directional vectors given by the projections of ${\bf V}_2/W_2$ 
and ${\bf V_3}/W_3$ which dominate the determination of $M_\star$ and 
$Y_0$ (cf.\ Fig.\ \ref{fig4}), and the ellipse is defined by these 
vectors.\label{fig5}}
\end{figure}

Inspecting the matrix ${\bf V}^{\rm T}$ in Figure \ref{fig4}, note that 
vector 2 reveals both $M_\star$ and $Y_0$ to be positive, indicating that 
they should be increased or decreased together to satisfy the constraints 
imposed by the observables. However, vector 3 implies that if $M_\star$ 
increases, $Y_0$ should decrease. When both of these vectors are projected 
onto the $M_\star$-$Y_0$ plane and scaled according to 1/$W_2$ and 
1/$W_3$, vector 3 is the longer of the two. These scaled vectors are shown 
as arrows in Figure~\ref{fig5}, defining the semi-major and minor axes of 
the projected 1$\sigma$ error ellipse. The error ellipse defines the 
boundary of the allowed parameter values that will change the $\chi^2$ 
value by less than $\Delta\chi^2$, where $\Delta\chi^2$ is the value that 
defines the 1$\sigma$ error for the relevant number of degrees of freedom. 
This figure shows clearly that there is a very strong inverse correlation 
between these parameters, and hints at the size of the uncertainty in each 
one\footnote{The correlation arises naturally from the fact that 
increasing either $M_\star$ or $Y_0$ tends to increase the luminosity, so 
if one of these parameters is increased the other must be decreased to 
satisfy the luminosity contraint while continuing to match the other 
observables.}. Though not as clearly visible in Figure \ref{fig4}, $Z$ and 
$Y_0$ have a strong direct correlation, so for the local analysis we 
explore a range of $Y_0$ as well as $Z$.

Finally, the {\it covariance matrix} can be expressed in a very compact 
form using SVD; the covariance between parameters $P_j$ and $P_l$ is:
\begin{equation}
C_{jl} = \sum_{k=1}^N \frac{V_{jk}V_{lk}}{W_{kk}^2}.
\label{eqn:covariance}
\end{equation}
In particular, the formal uncertainties $\sigma(P_j)$ on each of the 
parameters are obtained from $\sigma(P_j)^2 = \sum_k V_{jk}^2/W_{kk}^2$. 
Given this very clear form for describing the correlations between the 
parameters and their uncertainties, we can also determine the 
uncertainties on the model observables (such as $R_{\star}$ and 
$L_{\star}$) or some other properties (such as the core hydrogen content) 
by calculating the models at the boundaries of the error ellipse. 
Equation~(\ref{eqn:covariance}) can also be used to investigate {\it how} 
to reduce the uncertainties in each of the parameters---with what 
precision do we need to measure the individual observables? What other 
information may be useful to help break the degeneracies between some 
parameters? And are there limits to what we can learn from the available 
data?


\end{document}